\documentclass[conference]{IEEEtran}
\IEEEoverridecommandlockouts
\usepackage{makeidx}  
\usepackage{amssymb}
\usepackage{color}
\usepackage{graphicx}
\usepackage{subfigure}
\usepackage{url}
\usepackage{makecell}
\usepackage{cite}
\usepackage{listings}
\usepackage{paralist}
\let\subparagraph\paragraph
\usepackage{cite}
\usepackage{amsfonts, amsmath, amssymb}
\usepackage{epstopdf}
\usepackage{float}
\usepackage{enumerate}
\usepackage{color}
\usepackage{multirow}
\usepackage{verbatim}
\usepackage{stfloats}
\usepackage{xcolor}
\usepackage[utf8]{inputenc}
\usepackage{amsfonts}
\usepackage{eufrak}
\usepackage{eucal}
\usepackage{amsmath}
\usepackage{makeidx}
\usepackage{bm}

\makeatletter
\newif\if@restonecol
\makeatother

\usepackage[ruled,vlined, linesnumbered]{algorithm2e}
\makeatletter
\def\footnoterule{\kern-3\p@
  \hrule \@width 2in \kern 2.0\p@} 
\makeatother

\makeatletter
\def\BState{\State\hskip-\ALG@thistlm}
\makeatother

\definecolor{codegreen}{rgb}{0,0.6,0}
\definecolor{codegray}{rgb}{0.5,0.5,0.5}
\definecolor{codepurple}{rgb}{0.58,0,0.82}
\definecolor{backcolour}{rgb}{0.95,0.95,0.92}

\definecolor{dkgreen}{rgb}{0,0.6,0}
\definecolor{gray}{rgb}{0.5,0.5,0.5}
\definecolor{mauve}{rgb}{0.58,0,0.82}
\lstdefinestyle{mystyle}{
  backgroundcolor=\color{backcolour},   commentstyle=\color{codegreen},
  keywordstyle=\color{magenta},
  numberstyle=\tiny\color{codegray},
  stringstyle=\color{codepurple},
  basicstyle=\footnotesize,
  breakatwhitespace=false,
  breaklines=true,
  keepspaces=true,
  numbers=left,
  numbersep=5pt,
  showspaces=false,
  showstringspaces=false,
  showtabs=false,
  tabsize=2
}

\newtheorem{definition}{Definition}

\makeatletter
\newcommand{\tabincell}[2]{\begin{tabular}{@{}#1@{}}#2\end{tabular}}
\newcommand{\xRightarrow}[2][]{\ext@arrow 0359\Rightarrowfill@{#1}{#2}}
\makeatother

\newcommand{\simu}{\textsc{Simulink}}
\newcommand{\staf}{\textsc{Stateflow}}

\newcommand{\sat}{\textsc{Sat}}
\newcommand{\smt}{\textsc{Smt}}

\begin{document}
%
\title{Formal Analysis of Hybrid-Dynamic Timing Behaviors in Cyber-Physical Systems}
\author{
\IEEEauthorblockN{
Li Huang\IEEEauthorrefmark{1} and
Eun-Young Kang\IEEEauthorrefmark{2}
 }
\IEEEauthorblockA{\IEEEauthorrefmark{1}School of Data \& Computer Science, Sun Yat-Sen University, Guangzhou, China \\
huangl223@mail2.sysu.edu.cn}
\IEEEauthorblockA{\IEEEauthorrefmark{2}The Maersk Mc-Kinney Moller Institute, University of Southern Denmark, Denmark\\
eyk@mmmi.sdu.dk}
}
\maketitle

%

\begin{abstract}
Ensuring correctness of timed behaviors in cyber-physical systems (CPS) using closed-loop verification is challenging due to the hybrid dynamics in both systems and environments.
\simu\ and \staf\ are tools for model-based design that support a variety of mechanisms for modeling and analyzing hybrid dynamics of real-time embedded systems. In this paper, we present an \smt-based approach for formal analysis of the hybrid-dynamic timing behaviors of CPS modeled in \simu\ blocks and \staf\ states (S/S). The hierarchically interconnected S/S are flattened and translated into the input language of \smt\ solver for
formal verification. A translation algorithm is provided to facilitate the translation. Formal verification of timing constraints against the S/S models is reduced to the validity checking of the resulting \smt\ encodings. The applicability of our approach is demonstrated on an unmanned surface vessel case study.
\end{abstract}

\section{Introduction}
\vspace{-0.05in}
Cyber-Physical Systems (CPS) are real-time embedded systems in which the software controllers continuously interact with physical environments.
The continuous timed behaviors of CPS often involve complex dynamics as well as stochastic characteristics.
Formal verification and validation (V\&V) technologies are indispensable and highly recommended for development of safe and reliable CPS \cite{iso26262,iec61508}.
Ensuring correctness of timing properties in CPS using closed-loop system verification is challenging due to the continuous timed behaviors (described using ordinary differential equations) as well as non-linear dynamics (e.g., trigonometric and exponential functions) in CPS.
\simu\ and \staf\ (S/S) \cite{slsf} are widely-used industrial tools for model-based design of real-time embedded systems, which provide a variety of mechanisms for modeling and analyzing hybrid dynamic behaviors. \simu\ is a block-diagram based formalism that models time-continuous dynamics while \staf\ specifies control logic and state-based system behaviors.
The latest release of S/S  provides a new type of formalism, called \emph{Simulink-based state} (\emph{sl-state}) \cite{sl-state}, for descriptions of hybrid dynamic behaviors. \emph{sl-state} allows to embed various \simu\ blocks inside \staf\ states. A \staf\ chart consisting of a set of \emph{sl-states} is a graphical hybrid automaton (GHA) \cite{rajhans2018graphical}.

One of the solution to enable formal analysis of S/S models is to transform the models into satisfiability modulo theory (\smt) \cite{de2009satisfiability} formulas and perform \emph{bounded model checking} using \smt\ solvers.
However, the hybrid dynamics of real-time CPS modeled in S/S usually involve differential equations and non-linear functions, making the satisfaction problems of these formulas undecidable.
To alleviate this problem,  the approach of $\delta$-complete decision procedures \cite{gao2012delta} is proposed to check the satisfiability of such formulas under a tolerable numerical perturbation $\delta>0$. \emph{dReal} \cite{gao2013dreal} is an \smt\ solver that implements the $\delta$-complete decision procedures, which is able to analyze real-valued non-linear functions.

In our earlier works \cite{sscps, setta18}, we have provided the approaches for formal analysis of timing constraints in CPS modeled in S/S using Simulink Design Verifier (SDV) \cite{SLDV}, which, however, are limited to discrete-time S/S models with no non-linear functions.
To enable the formal analysis of the continuous-time behaviors in CPS involved with non-linear dynamics,
in this paper, we propose an \smt-based approach for analyzing CPS modeled in GHA using \emph{dReal}:
\begin{inparaenum}
\item Formal definitions of \emph{Simulink-based state} and GHA are provided;
\item The hierarchical GHA is flattened and translated into the input language of \emph{dReal};
\item A translation algorithm is provided to facilitate the translation;
\item Formal analysis of timing constraints against GHA is reduced to the validity checking of the resulting \smt\ encodings.
\end{inparaenum}
Our approach is demonstrated on an unmanned surface vessel (USV) case study.

The paper is organized as follows: Sec. \ref{sec:preliminary} presents an overview of S/S, GHA and \emph{dReal}.
The translation from GHA to \smt\ language is presented in Sec. \ref{sec:translation}.  The
applicability of our approach is demonstrated by performing verification on the USV in Sec. \ref{sec:case-study}.
Sec. \ref{sec:related-work} and Sec. \ref{sec:conclusion} present related works and conclusion.
\vspace{-0.1in}
\section{Preliminary}

\label{sec:preliminary}
\noindent\textbf{Simulink and Stateflow (S/S)} \cite{slsf} is a synchronous data flow language that provides supports for modeling and simulation of real-time embedded systems, as well as code generation. An S/S model represents a diagram composed of various types of \emph{blocks} (e.g., {\tiny $\ll$}Sum{\tiny $\gg$}, {\tiny $\ll$}Product{\tiny $\gg$}) interconnected via lines and I/O ports. A hierarchical S/S model of an arbitrary depth can be achieved using composite blocks, i.e., \emph{subsystems}.
The recent version of S/S has provided a new type of formalism, called \emph{Simulink based state} (\emph{sl-state}), specialized for modeling hybrid  behaviors incorporating both discrete and time-continuous dynamics. An \emph{sl-state} has a subsystem embedded inside, which describes the dynamical behaviors when the \emph{sl-state} is active.
A \staf\ chart that consists of a set of \emph{sl-states} is called a graphical hybrid automaton (GHA) \cite{rajhans2018graphical}.

\vspace{0.05in}
\noindent\textbf{SMT and dReal Solver:}
Satisfability  (\sat) problem is the problem to determine whether a set of propositional formulas can be true by assigning true/false values to the constituent variables.
Satisfability Modulo Theories (\smt) \cite{de2009satisfiability} is an extension of \sat, in which the symbols of the input formulas are given with background theories (e.g., arithmetic, array). \smt-based bounded model checking is to check whether a given requirement $\phi$  (encoded as a logical formula) is valid over a model $\mathbb{M}$ (represented by a set of \smt\ formulas) by querying whether $\mathbb{M}\wedge \neg \phi$ is unsatisfiable.
\emph{dReal} \cite{gao2013dreal} is an \smt\ solver that employs the $\delta$-complete decision procedures \cite{gao2012delta}, which allows users to check satisfiability of formulas with non-linear real-valued functions.  Given a set of formulas, \emph{dReal} checks whether the formulas are satisfied up to a given precision $\delta$ (i.e., a user-defined parameter that represents the maximum tolerable numerical errors). The input language of \emph{dReal} follows SMT-LIB standard \cite{barrett2017smt} and includes extended formalisms expressing ordinary differential equations (ODEs) and non-linear functions.

\section{Translation of GHA into dReal}
\label{sec:translation}
In this section, we investigate how to translate GHA models into the input language of \emph{dReal}.
We first show how to flatten the hierarchical \emph{sl-states} in GHA and then give the formal definitions of the \emph{sl-state} and GHA, followed by brief descriptions of a translation algorithm.

\begin{figure}[htbp]
  \centering
  \includegraphics[width=3.5in]{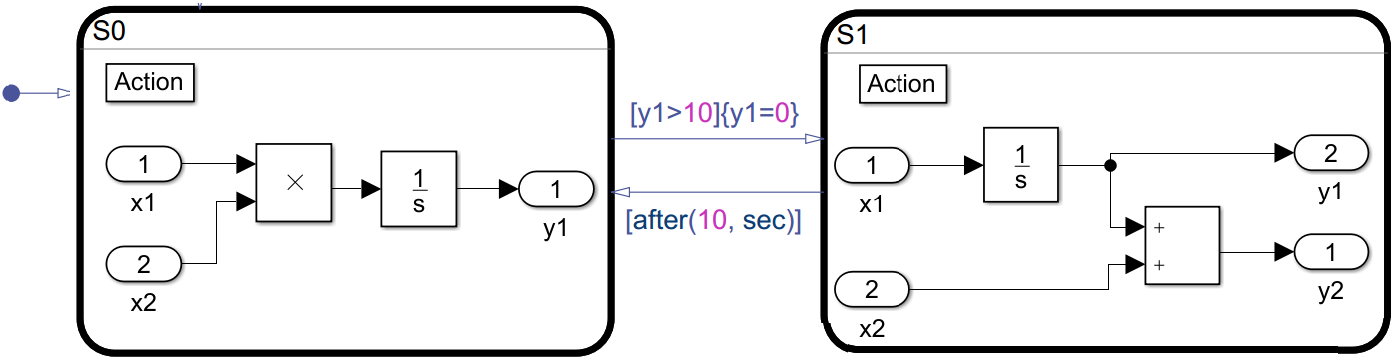}
  \caption{{\scriptsize An example of GHA: 1) when $S0$ is active: $y1=\int(x1+x2)$;
2) when the transition $S0\rightarrow S1$ is taken: $y1=0$;
3) when $S1$ is active: $y1=\int x1$, $y2=x2+\int x1$.}
}
  \label{fig:example}
\end{figure}

In GHA, a hierarchical \emph{sl-state} is achieved by using \emph{subsystems}, i.e., composite blocks that encapsulate a set of atomic \simu\ blocks and possibly other \emph{subsystems}. A \emph{subsystem} can be flattened by replacing it with the set of interconnected blocks originally embedded in it. By recursively flattening the \emph{subsystems} at arbitrary nesting levels, the hierarchy of \emph{sl-state} can be eliminated.
An example of GHA is shown in Fig. \ref{fig:example}, in which the behaviors of the two \emph{sl-states} $S0$ and $S1$ are described by a set of \simu\ blocks (e.g., {\tiny $\ll$}Product{\tiny $\gg$},  {\tiny $\ll$}Integrator{\tiny $\gg$}) embedded in the states.

\vspace{-0.1in}
\vspace{-0.1in}

To present the mapping from GHA to \smt\ formulas, the syntactic definitions of \emph{sl-state} and GHA are provided.
\vspace{0.05in}
\begin{definition}
\textbf{\emph{(Simulink-based State ({sl-state}))}} An \emph{sl-state} is a tuple $\langle\nu,\ B,\ L\rangle$, where
\begin{itemize}
\item $\nu$ is a finite set of variables (or signals);
\item $B$ is a finite set of atomic blocks that represent various mathematical/logical functions. A block $b\in B$ is a tuple
$\langle v_{in}$, $v_{out}$, $p$, $f\rangle$, where $v_{in}$, $v_{out}$ are the input/output variables of $b$. $p$ is a set of internal parameters. $f$ represents the update function that changes the value of $v_{out}$ based on  $v_{in}$ and $p$, i.e,  $f: v_{in} \times p \mapsto v_{out}$;
\item $L$ is a finite set of lines that connect blocks in $B$. A line $l$ propagates the data flow of a variable from the source block to the destination block(s).
\end{itemize}
\end{definition}

\vspace{0.05in}
\begin{definition}
\textbf{\emph{(Graphical Hybrid Automaton (GHA))}} A GHA is a transition system $\langle V_{in},\ V_{out},\ S,\ T \rangle$, where
\begin{itemize}
\item $V_{in}$ and $V_{out}$ are the sets of input/output variables;
\item $S$ is a set of \emph{sl-states};
\item $T$ is a set of transitions between different \emph{sl-states} in $S$. A transition $t\in T$ is a tuple $\langle\nu$, $src$, $dst$, $cond$, $act \rangle$, where
\begin{itemize}
\item[--] $\nu$ is a set of variables on the transition;
\item[--] $src$, $dst\in S$ are the source/destination states of the transition;
\item[--] $cond$ is the condition (i.e., a propositional formula or a temporal logic expression) that should be evaluated as true when the transition is taken;
\item[--] $act$ represents a set of actions executed when the transition is activated.
\end{itemize}
\end{itemize}
\end{definition}

Let $\mathbb{M}= \langle V_{in},\ V_{out},\ S,\ T \rangle$ be a GHA. $\forall i\in \mathbb{N}, \ 1\leq i\leq k$, $s_i$ denotes the active state of $\mathbb{M}$ at the $i^{th}$ step and $t_i$ represents the $i^{th}$ transition (i.e., transition from $s_i$ to $s_{i+1}$). An execution of $\mathbb{M}$ with $k$ steps (i.e., $k$ transitions) can be defined as the following sequence of states:
\[s_0 \xrightarrow{t_0} s_1 \xrightarrow{t_1} \cdot\cdot\cdot s_i \xrightarrow{t_i} s_{i+1}\xrightarrow{t_{i+1}} \cdot\cdot\cdot  s_k\]

Let $v \in V_{out}$ be an output variable of $\mathbb{M}$. For each step $i$, two variables $v_{i}^{\circ}$ and $v_{i}^{\ast}$ are utilized to denote the values of $v$ at the beginning and end of the $i^{th}$ step, respectively.
The variation of $v$ at the $i^{th}$ step can be represented by the relation between $v_{i}^{\circ}$ and $v_{i}^{\ast}$:
(1) $v\notin s_i.\nu \Longrightarrow v_{i}^{\ast} = v_{i}^{\circ}$;
(2) $v\in s_i.\nu \Longrightarrow v_{i}^{\ast} = \Phi_{s_i}^{v}$.
Here $\Phi_{s_i}^{v}$ is the formula that represents the functions or operations updating the value of $v$ when $\mathbb{M}$ stays in the state $s_i$.
Similarly, the variation of $v$ at transition $t_i$ can be described as the relation between $v_{i}^{\ast}$ and $v_{i+1}^{\circ}$:
(1) $v \notin t_i.\nu \Longrightarrow v_{i+1}^{\circ} = v_i^{\ast}$;
(2) $v \in t_i.\nu \Longrightarrow v_{i+1}^{\circ} = \Pi_{t_i}^{v}$. Here, $\Pi_{t_i}^{v}$ is the action that updates the value of $v$ when the transition $t_i$ is taken.

We call the formulas $\Phi_{s_i}^{v}$ and $\Pi_{t_i}^{v}$ the \emph{formula representation} (FR) of variable $v$ in state $s_i$ and at transition $t_i$, respectively. For instance, in the GHA depicted in Fig. \ref{fig:example}, the FR of the output variable $y2$ in state $S1$ is $\Phi_{S1}^{y2} = x2+\int x1$.

To perform the bounded model checking on the GHA model using \emph{dReal}, we first derive the FRs of the output variables of $\mathbb{M}$  and then specify the relations between the variables and the corresponding FR at each step using \smt\ assertions.

The FRs of the output variables in an \emph{sl-state} can be derived based on the dynamic behaviors of the \emph{sl-state}, described by a set of embedded \simu\ blocks. These blocks include computational blocks (e.g., {\tiny $\ll$}Sum{\tiny $\gg$}, {\tiny $\ll$}Product{\tiny $\gg$}) and time-continuous blocks (e.g., {\tiny $\ll$}Integrator{\tiny $\gg$}).
The FR of the output variable of an atomic computational block can be derived based on its inputs, parameters and transfer function.
In \emph{dReal}, the continuous-time integral functions (i.e., ODEs) are defined using \emph{flow declarations} and \emph{flow conditions}, which specify the derivatives and inputs/outputs of the integrals.
To represent the semantics of the {\tiny $\ll$}Integrator{\tiny $\gg$} blocks in \emph{dReal},  the derivatives of the integrals are define using  \emph{flow declarations}, and the relations between the input and output variables of the integrals can be specified by \emph{flow conditions}.
After the FRs of the output variables of all atomic blocks are derived, the FRs of the output variables of \emph{sl-state} can be derived.
Given an output variable $v$ in an \emph{sl-state} $s$, the FR of $v$ in $s$ (represented by $\Phi_{s}^{v}$) can be deduced by recursively replacing the constituent variables of $\Phi_{s_i}^{v}$ with their corresponding FRs.

Based on the FRs of the output variables of GHA, the behaviors of GHA can be represented in \emph{dReal} using \smt\ assertions. The translation from GHA into \smt\ encodings can be summarized in Algorithm \ref{alg1}, in which $V_{cin}$ ($V_{cout}$) denotes a set of input (output) variables of {\tiny $\ll$}Integrator{\tiny $\gg$} blocks in GHA.
The \smt\ encodings that contain the $k$-step behaviors of GHA can be generated by the following steps:

\noindent 1. For each variable $v$, add the assertions that describe the relation between the values of $v$ at the beginning and at the end of each step (line 4--9).
If $v$ is an output variable of an integral, then $v$ is updated based on the corresponding ODE (denoted $fl_s$). Otherwise, $v$ can  either be updated based on the FR of $v$ in the current state or remain unchanged.

\noindent 2. Add the \smt\ assertions to specify the behaviors of each transition and the changes of values of variables involved in the transitions (line 11--14).

\begin{algorithm}[htbp]
\small
    \caption{{\small{Translation from GHA to \smt\ language}}}
    \label{alg1}
    \KwIn{GHA model represented by $\langle V_{in},\ V_{out},\ V_{cin},\ V_{cout},\ S,\ T\rangle$, verification bound $k$} \KwOut{\smt\ encodings}
    {
    \For{i=0, i$<$k, i++}{ 
    \For{each $v$ in $V_{out} \bigcup V_{cin} \bigcup V_{cout} $}{
    \For{each $s\in S$}{
    \If{$v\in V_{cout} \wedge v\in s.\nu$}{ 
    $Assert(s_i=s\implies(v_{i}^{\ast} = v_{i}^{\circ} + \int fl_{s}))$
    }
    \ElseIf{$v\notin V_{cout} \wedge v \in s.\nu$}{

    $Assert(s_i=s \implies v_{i}^{\ast} =\Phi_{v}^{s_i})$ \\

    }
    \Else{
    $Assert(s_i=s \implies v_{i}^{\ast} = v_{i}^{\circ} )$ \\ 

    }
    \vspace{-0.03in}
    }
    \For{each $t\in$ $T$}{
    \If{$v\in t.\nu$}{ 
    $Assert((s_i=t.src \wedge t.cond) \implies (s_{i+1}=t.dst \wedge  v_{i+1}^{\circ} = \Pi_v^{t}))$
    }
    \Else{ 
    $Assert((s_i=t.src \wedge t.cond) \implies (s_{i+1}=t.dst \wedge  v_{i+1}^{\circ} = v_{i}^{\ast}))$
    }
    \vspace{-0.03in}
    }

    \vspace{-0.03in}
    }
     \vspace{-0.02in}
    }
    \vspace{-0.05in}
    }

\end{algorithm}

\vspace{-0.05in}
\section{Case Study}
\label{sec:case-study}
Our approach is demonstrated on an unmanned surface vessel (USV) \cite{usv}, which is a propeller-driven vessel that can automatically adjust its movements to reach the user-defined target location.
USV can replace human operators to transport to hard-to-reach places or hazardous sites, which avoids to expose humans to risks and reduces operational costs.
Equipped with four sensors (i.e., GPS, compass, speed sensor and accelerometer), USV is able to obtain the information of its  position, heading direction, speed and acceleration. USV is capable of changing its position (represented ``(x, y)'') by
moving forward/backward (surge) or left/right (sway) in two perpendicular axes. The orientation (heading direction) of USV can be changed through rotation. A set of representative timing constraints on USV are listed below.\\
\noindent R1.	USV can finally arrive at the target location and stop within a given time, e.g., 20 seconds.\\
\noindent R2.	If the heading error exists, i.e., the heading direction is different with the desired direction to the target location, USV should decelerate and adjust the heading direction within a certain time, e.g., 500ms.\\
\noindent R3.	The GPS component should capture the position of USV periodically with period 50 ms and jitter 10ms. \\
\noindent R4.	The accelerometer component should be triggered every [20, 30]ms, i.e., a periodic acquisition of accelerometer must be carried out to update the acceleration of USV.

We construct a GHA model that describes the hybrid-dynamic behaviors of USV, which consists of four \emph{sl-states} with 188 blocks distributed in the \emph{sl-states}.
The stochastic behaviors of the physical environments (i.e., unpredictable speeds of wind and current) are
represented by a set of pre-generated random parameters, which remain unchanged after initialized.
To enable the formal analysis of the continuous-time behaviors in USV with non-linear functions, we translate the USV GHA model into \smt\ formulas.
According to the translation process described in the previous section, we first extract the information of the relations between inputs and output variables of the USV GHA model, as well as the formulas that represent integral functions.
Based on the extracted information/formulas, assertions that describe those relations and functions are generated. Bounded model checking of the timing constraints on the resulting \smt\ encodings is perform by \emph{dReal}. The target location is configured as (50, 50). The verification results are illustrated in Table \ref{verification_result}.

\begin{table}[htbp]
\scriptsize
  \centering
   \renewcommand\arraystretch{1.3}
  \caption{Verification Results of USV in \emph{dReal} ($k=20$, $\delta=0.001$)}
    \begin{tabular}{|c|c|c|c|}
    \hline
    Req  & SMT   &\tabincell{l}{Time\\ (Min)} \\
    \hline
     R1 & s$_{20}$=stop$\wedge$(50$-$x$_{20}^{\ast}\leq$ 0.8)$\wedge$(50$-$y$_{20}^{\ast}\leq$0.8)$\wedge runT \leq 20$ & 89.46
         \\ \hline

     R2 & $\bigwedge\limits_{i=1}^{20}$(hError$_i^{\ast}>$0.01$\implies$dec$_{i+1}^{\circ}$=true$\wedge\ reactT \leq 0.5$) & 415.68
         \\ \hline

     R3 & $\bigwedge\limits_{i=1}^{20}$ \{(gps\_t$_{i+1}-$gps\_t$_i\geq$0.04)$\wedge$(gps\_t$_{i+1}-$gps\_t$_i\leq$0.06)\}  & 57.16
         \\ \hline
     R4 & $\bigwedge\limits_{i=1}^{20}$ \{(acc\_t$_{i+1}-$acc\_t$_i\geq$0.02)$\wedge$(acc\_t$_{i+1}-$acc\_t$_i\leq$0.03)\}  & 49.30
    \\ \hline
    \end{tabular}%
  \label{verification_result}%
\end{table}%
\vspace{-0.1in}

\section{Related Work}

\label{sec:related-work}

Considerable efforts have been made in the integration of S/S and formal techniques for analysis of CPS.
Approaches for formal analysis of CPS behaviors modeled in S/S using the integrated tool SDV
 were investigated in several works \cite{setta18, sac18, JFSLDV}, which are only applicable for verification of discrete-time S/S models constructed by a restricted set of \simu\ blocks. To perform formal analysis using other tools, transformations of S/S models into other verifiable formalisms have been investigated in the earlier studies \cite{tripakis2005translating, minopoli2016sl2sx, reicherdt2014formal}.
Minopoli et al. \cite{minopoli2016sl2sx} translated S/S models into the formats recognized by the verification platform SpaceEx and Tripakis et al. \cite{tripakis2005translating} translated S/S into the synchronous data flow language Lustre. Reicherdt et al. \cite{reicherdt2014formal} presented a theorem-proving approach by translating S/S models into the input language of Boogie tool. However, in those works, neither formal definitions of S/S (with respect to the updated modeling formalisms) nor analysis support for S/S models containing non-linear hybrid dynamics were provided.
Robert et al. \cite{herber2013bit} presented an \smt-based analysis approach via transformation from S/S into the input language of UCLID
verifier. However, the non-linear functions were over-approximated during the translation, which reduces the preciseness of the approach.
Filipovikj et al. \cite{filipovikj2019bounded} generated execution paths of hybrid system models in S/S and encoded the paths into \smt\ formulas amenable to formal analysis using Z3 solver, whereas, our work focuses on enabling formal analysis of hybrid system behaviors modeled in GHA formalism and provides the approach to translate GHA into  \smt\ formulas recognized by \emph{dReal} solver.

\section{Conclusion}

\label{sec:conclusion}
We present an \smt-based approach for formal analysis of hybrid-dynamic timing behaviors in cyber-physical systems (CPS) modeled as graphical hybrid automaton (GHA).
To enable formal analysis of GHA, the formal definitions of \emph{Simulink-based state} and GHA are provided.
The hierarchical GHA is flattened and translated into formulas amenable to \smt\ solving.
The formal verification of (non)-functional requirements of CPS is reduced to validity checking of the translated \smt\ formulas using \emph{dReal} solver. Moreover, a translation algorithm is provided to facilitate the translation. The feasibility of our approach is demonstrated by performing formal verification on an unmanned surface vessel case study.

Although, we have shown that translating GHA models into \emph{dReal} formula is sufficient to verify continuous timing behaviors involved with non-linear dynamics, the computational cost of the verification in terms of time is rather expensive. Thus, we continuously investigate complexity-reducing design for CPS to improve effectiveness and scalability of system design and verification.
As our ongoing work, we plan to develop the tool support for automatic translation. Furthermore, to formally analyze stochastic environmental behaviors in CPS, representing those behaviors CPS using advanced mathematical formalisms (e.g., stochastic differential equations) and enabling \smt-based analysis of those formalisms are further investigated.

\vspace{0.1in}
\noindent\textbf{Acknowledgment.} This work is supported by the EASY project funded by NSFC,
a collaborative research between Sun Yat-Sen University and University of Southern
Denmark.

\bibliographystyle{IEEEtran}
\bibliography{reference}

\end{document}